\documentclass{iaus}
\usepackage{graphicx}

\title[Galaxy evolution in the Cen A group]{Evolution of dwarf galaxies in the Centaurus A group}

\author[Makarova L., Makarov D.]{Makarova L.$^1$, Makarov D.$^1$}

\affiliation{$^1$Special Astrophysical Observatory, 
Nizhnij Arkhyz, Karachai-Cherkessian Republic, Russia 369167 \break email: lidia@sao.ru}

\pubyear{2007}
\volume{244}  
\jname{Dark Galaxies \& Lost Baryons}
\editors{A.C. Editor, B.D. Editor \& C.E. Editor, eds.}
\begin{document}

\maketitle

\begin{abstract}
We consider star formation properties of dwarf galaxies in Cen~A group observed within our HST/ACS projects number 9771 and 10235. 
We model color-magnitude diagrams of the galaxies under
consideration and measure star formation rate and metallicity dependence on time. 
We study environmental dependence of the galaxy evolution and
probable origin of the dwarf galaxies in the group.
\end{abstract}


\section{Introduction}
Centaurus A is one of the nearest group of galaxies at the mean distance
of 3.77 Mpc from us. The Cen~A/M~83 region is particularly interesting 
because one of the dominant galaxies Cen~A is a giant elliptical. 
The only other nearby large elliptical galaxy is the highly obscured Maffei~1.
The Cen~A group is probably more dynamically evolved than others in the Local Volume, like Local group or M~81 group, which have a spiral galaxy as central body.
At the distance of Cen~A group galaxies are well resolved into individual stars including old giant branch stars up to an age of 10 to 13 Gyr. The Centaurus A group contain 31 members or probable members (Karachentsev 2005, AJ 129, 178). 

\section{The dataset and the method}
Over the last several years members of our team participated in two snapshot surveys of nearby galaxies using WFPC2 aboard HST, which have provided us with the material for distances of $\sim 10 \%$ accuracy for about 150 nearby galaxies (proj. 8192 and 8601). Further significant progress has been made with HST/ACS pointed observations (proj. 9771 and 10235). Most of known dwarf members of the Cen~A/M~83 complex were imaged within the framework of these projects. Here we present star formation history (SFH) determination from color-magnitude diagrams (CMDs) of resolved stars in 14 dwarf galaxies in the Cen~A group (see Fig.1). The galaxies span absolute B magnitude range of $[-10\div-16]$. HST/ACS images of these objects were taken with F606W (broadband V) and F814W (broadband I) filters.  Photometry of resolved stars in the galaxies was made with the DOLPHOT package for crowded field photometry (Dolphin 2000, PASP 112, 1383). Photometric distances for all galaxies in the sample were obtained using Tip of the Red Giant Branch distance indicator (Karachentsev et al. 2007, AJ 133, 504).

\begin{figure}
\includegraphics{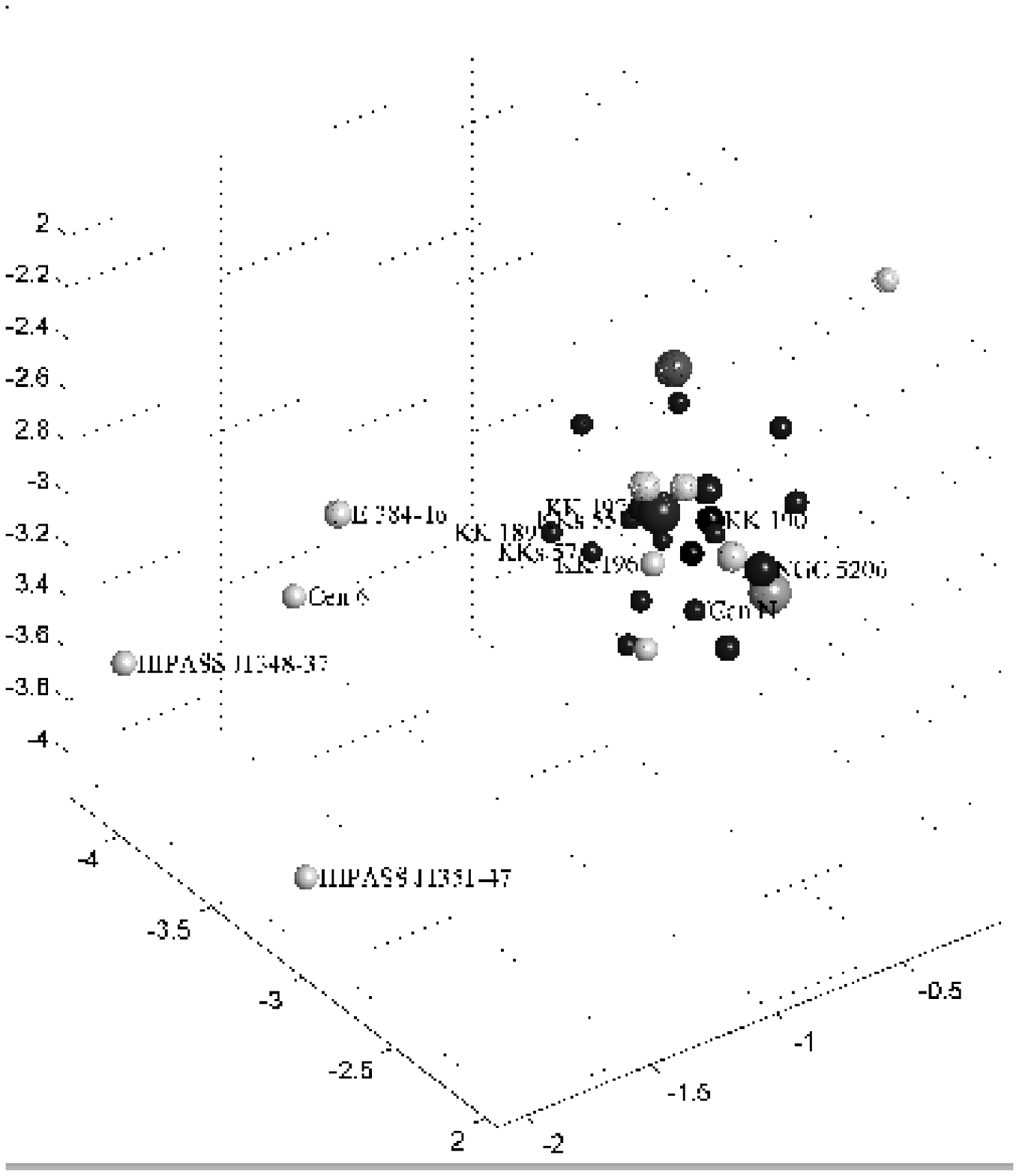}
\caption{3D structure of the Cen~A group is shown. Size of the spheres corresponds to absolute magnitude of a galaxy. Grey spheres represent dIrr galaxies and black spheres are dSphs. Three largest spheres are central body of the group (Cen~A elliptical galaxy) and two spiral members. Names of the galaxies with SFH measured by us are indicated in the figure. Well known morphological segregation between dSph and dIrr galaxies can be recognize in the picture. The dwarf spheroidals are closer to the central body, whereas the dwarf irregulars mostly situated in the outer part of the group.}
\end{figure}

We have created a program to analyze our large and homogeneous sample of galaxies (Makarov \& Makarova 2004, Ap 47, 229). We construct synthetic CMDs from theoretical stellar isochrones taking into account the initial mass function, galaxy distance, Galactic extinction and photometric errors. Photometric uncertainties and completeness values were added using results of artificial star tests. A linear combination of synthetic CMDs of different ages and metallicities forms a model CMD. For SFH determination we have to find a best linear combination of partial model CMDs to match the observed data. We construct a maximum-likelihood function for this task.

\begin{figure}
\hspace{-10mm}
\begin{tabular}{cc}
\includegraphics[width=0.55\textwidth]{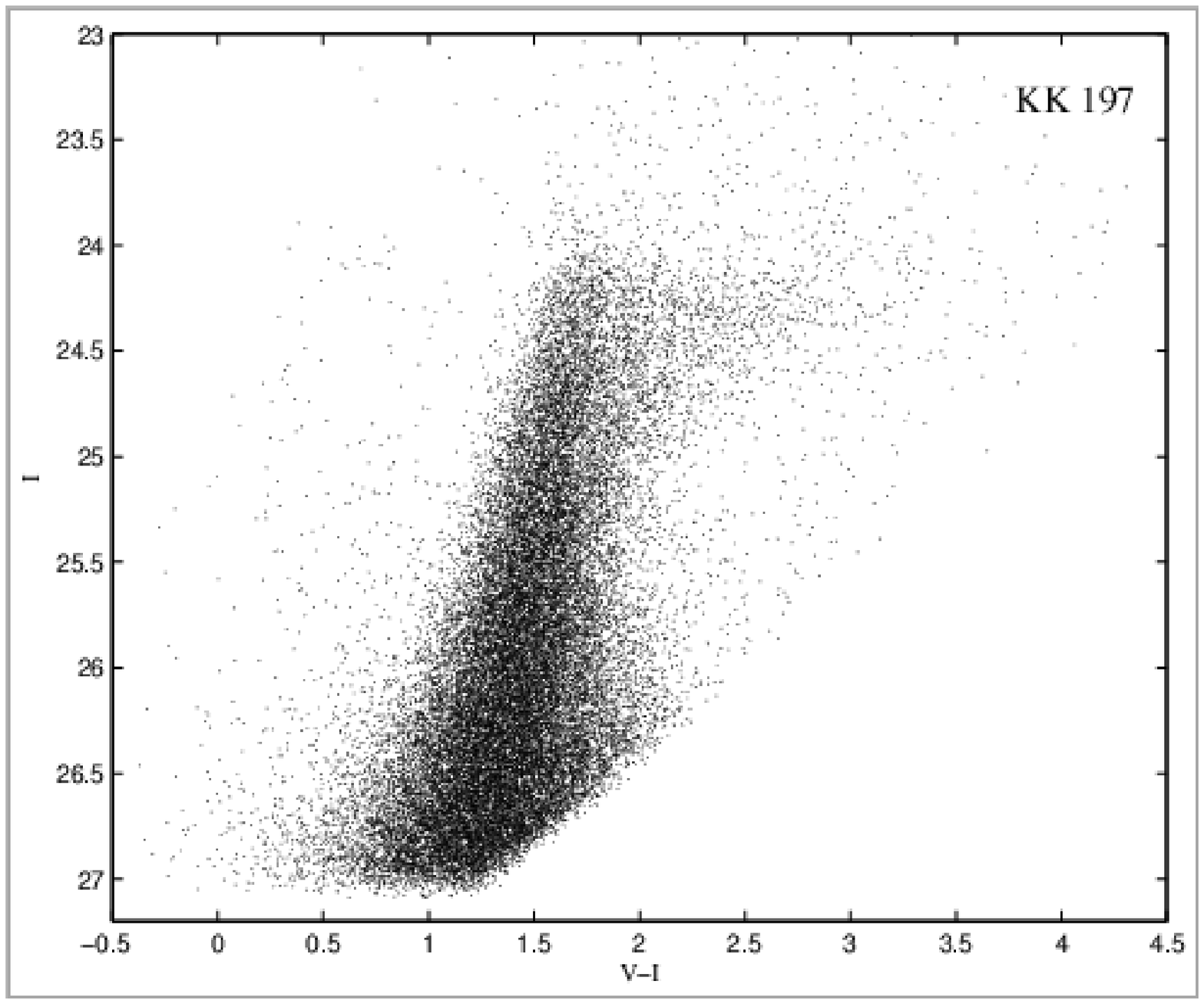}  &
\includegraphics[width=0.55\textwidth]{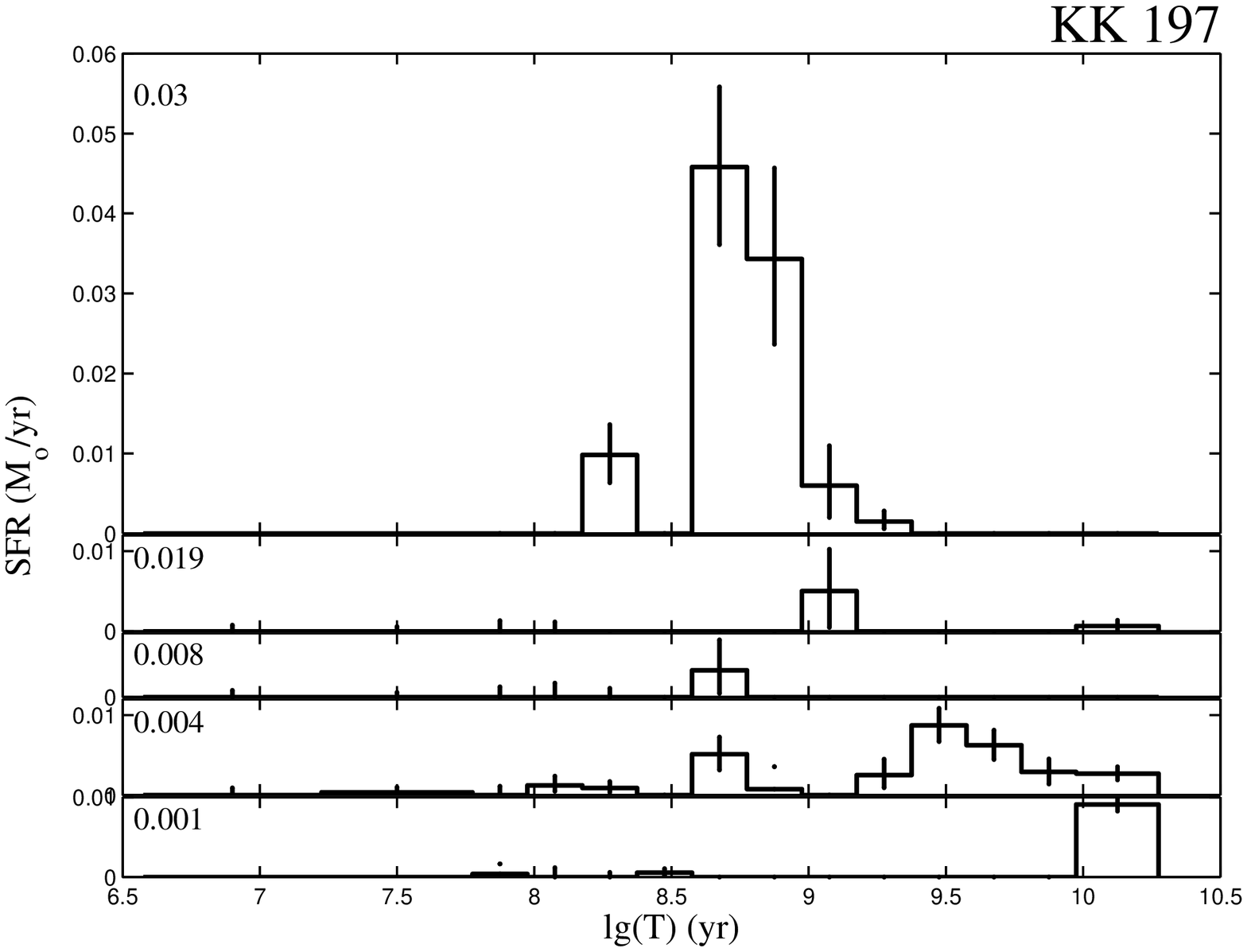}  \\
\includegraphics[width=0.55\textwidth]{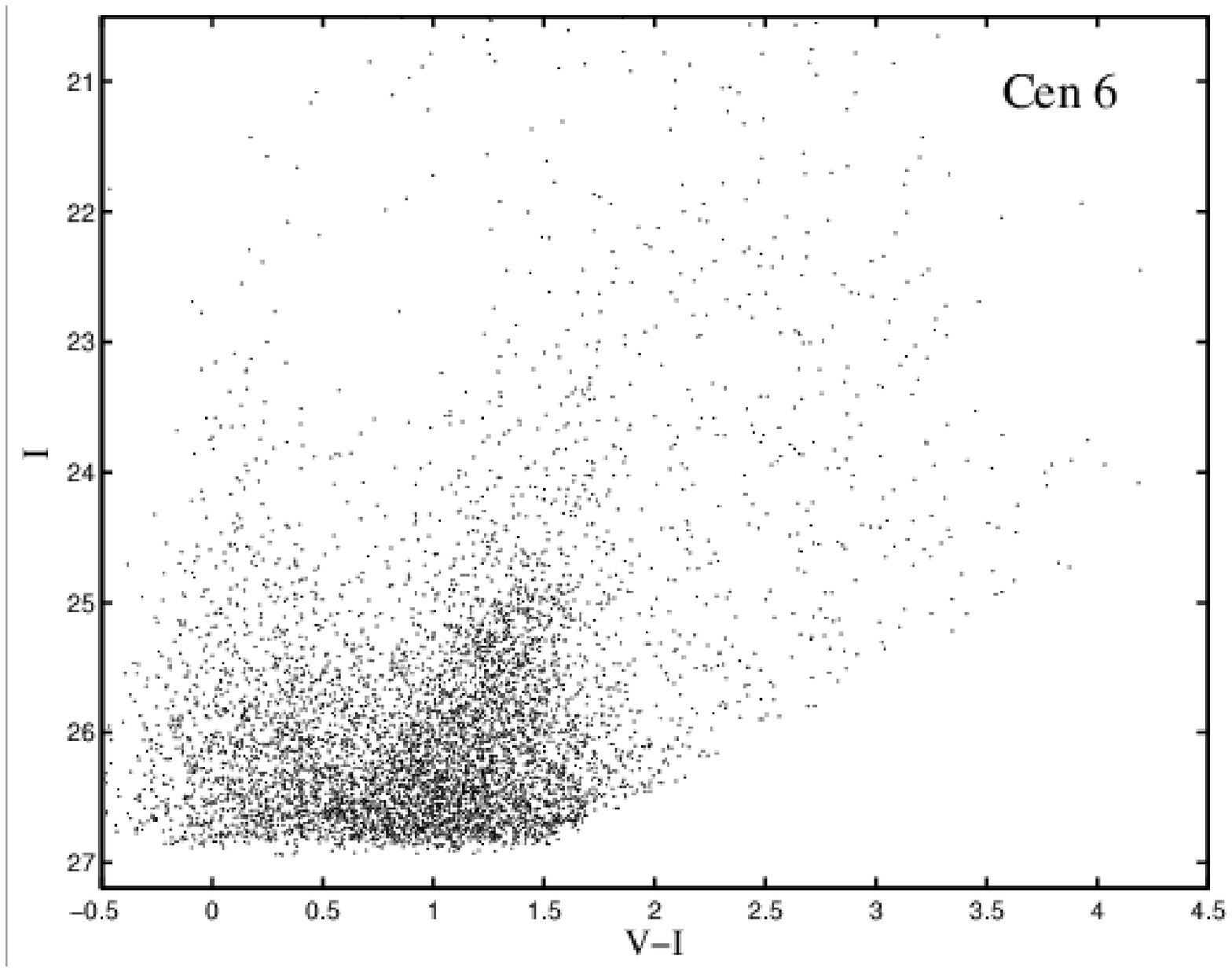}  &
\includegraphics[width=0.55\textwidth]{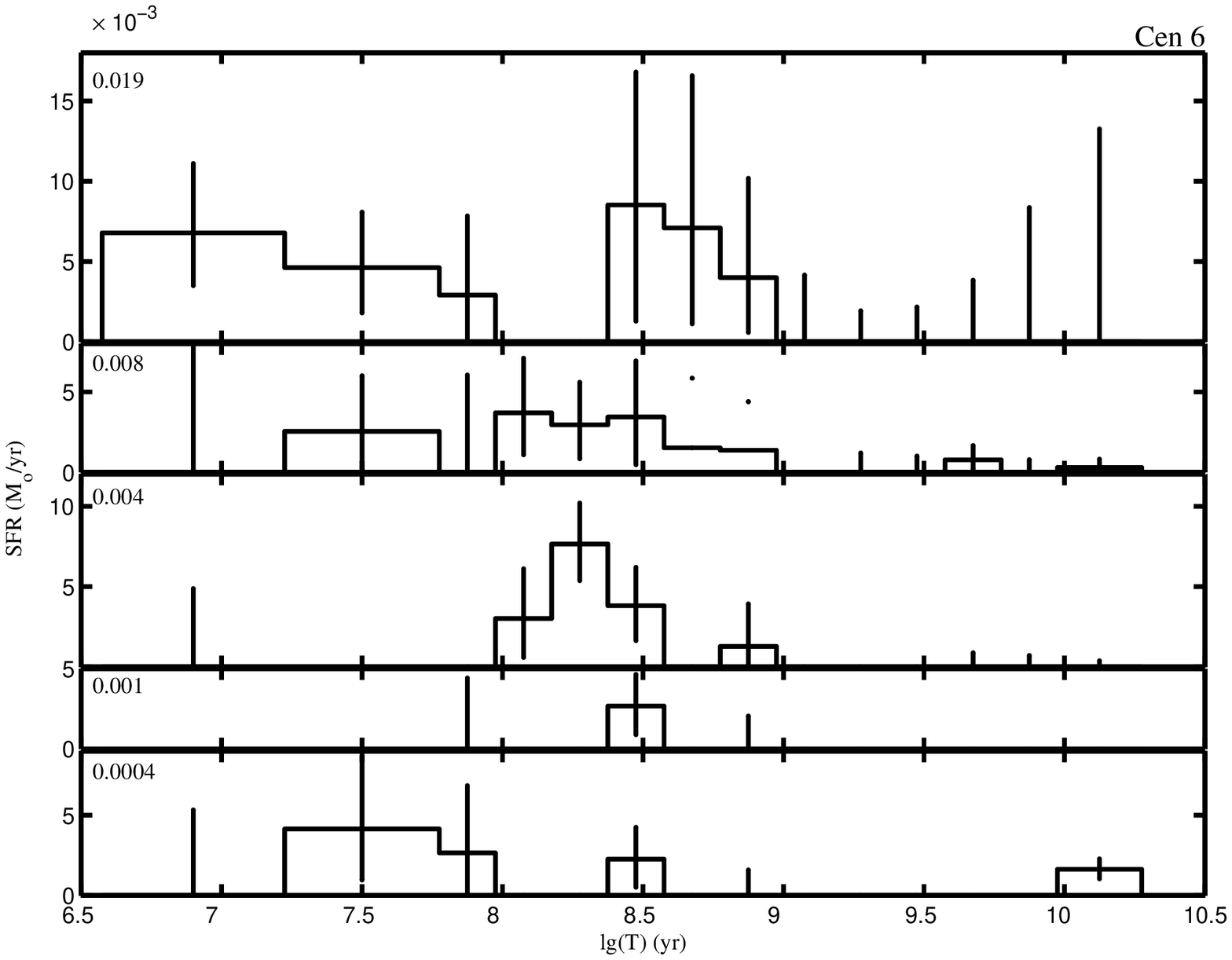}  
\end{tabular}
\caption{I, V$-$I CMDs of two dwarf galaxies in the Cen~A group (left panels). Photometric distance of KK~197 is 3.9 Mpc and the distance of Cen~6 is 5.8 Mpc. Star formation histories of KK~197 and Cen~6 are shown at the right panels.}
\end{figure}

\section{Results}
Color-magnitude diagrams and star formation histories of two dwarf galaxies in the Cen~A group are shown in Fig.2 to represent our sample. The dIrr galaxy Cen~6 is situated on the outskirts of the Cen~A group. We can see sparsely populated blue and red supergiant branches, AGB stars and upper red giant branch at the CMD. At least two star formation episodes can be distinguished (right panel): the ancient one, about 12--13 Gyr ago with low metallicity and more recent about 10--800 Myr ago with the metallicity probably up to solar.
The CMD of the dSph galaxy KK~197 contain well populated old RGB and some intermediate age AGB stars. The galaxy has unusual RGB color scatter, which reflects active star formation episode of very high metallicity level about 0.4--1 Gyr ago, whereas ancient burst of star formation has low metallicity.
It is worth to note, that only brightest part of stellar population of the galaxies can be seen, that puts some restrictions to details of our modeling.


\begin{figure}[p]
\begin{tabular}{cc}
\includegraphics[width=0.5\textwidth]{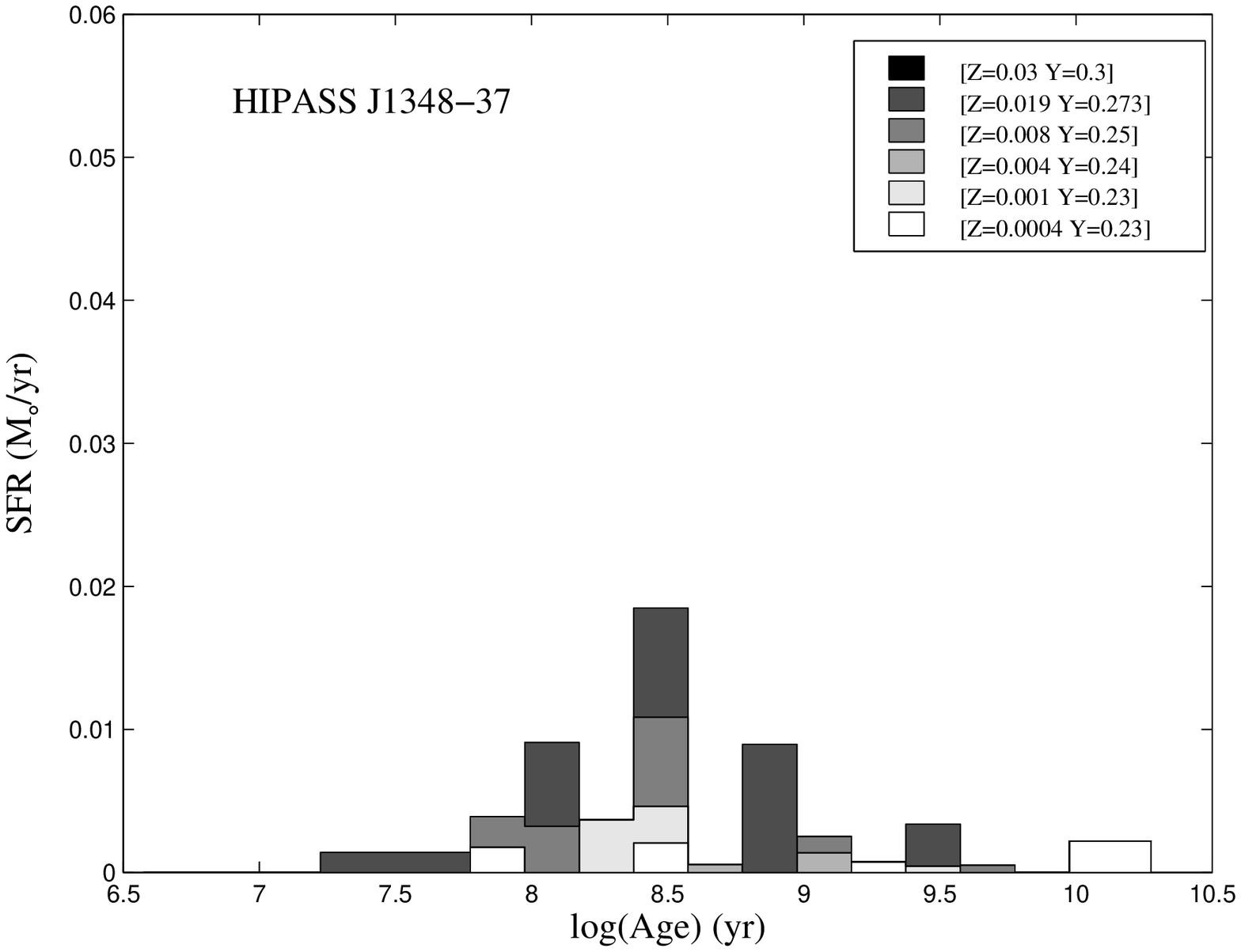}  &
\includegraphics[width=0.5\textwidth]{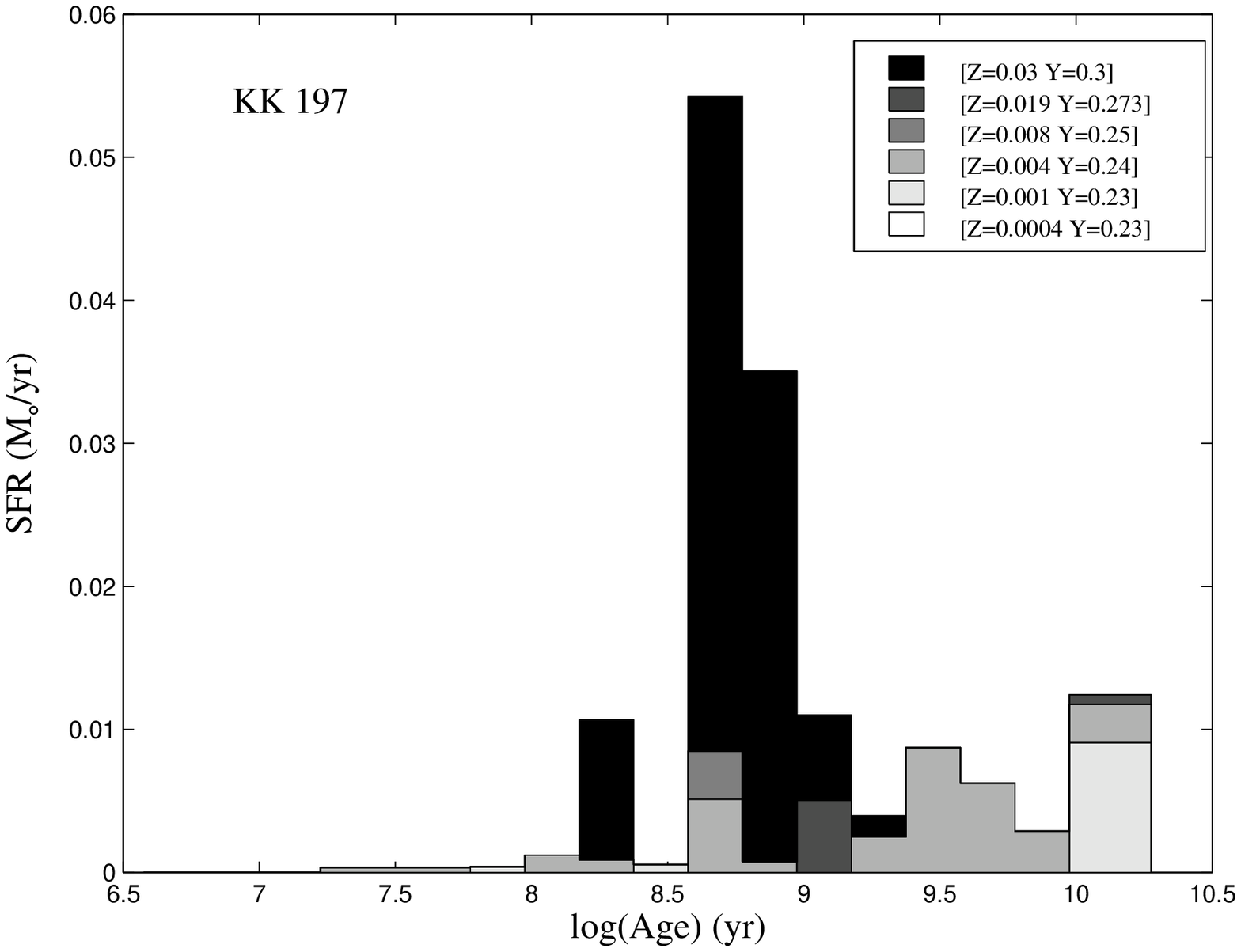}  \\
\includegraphics[width=0.5\textwidth]{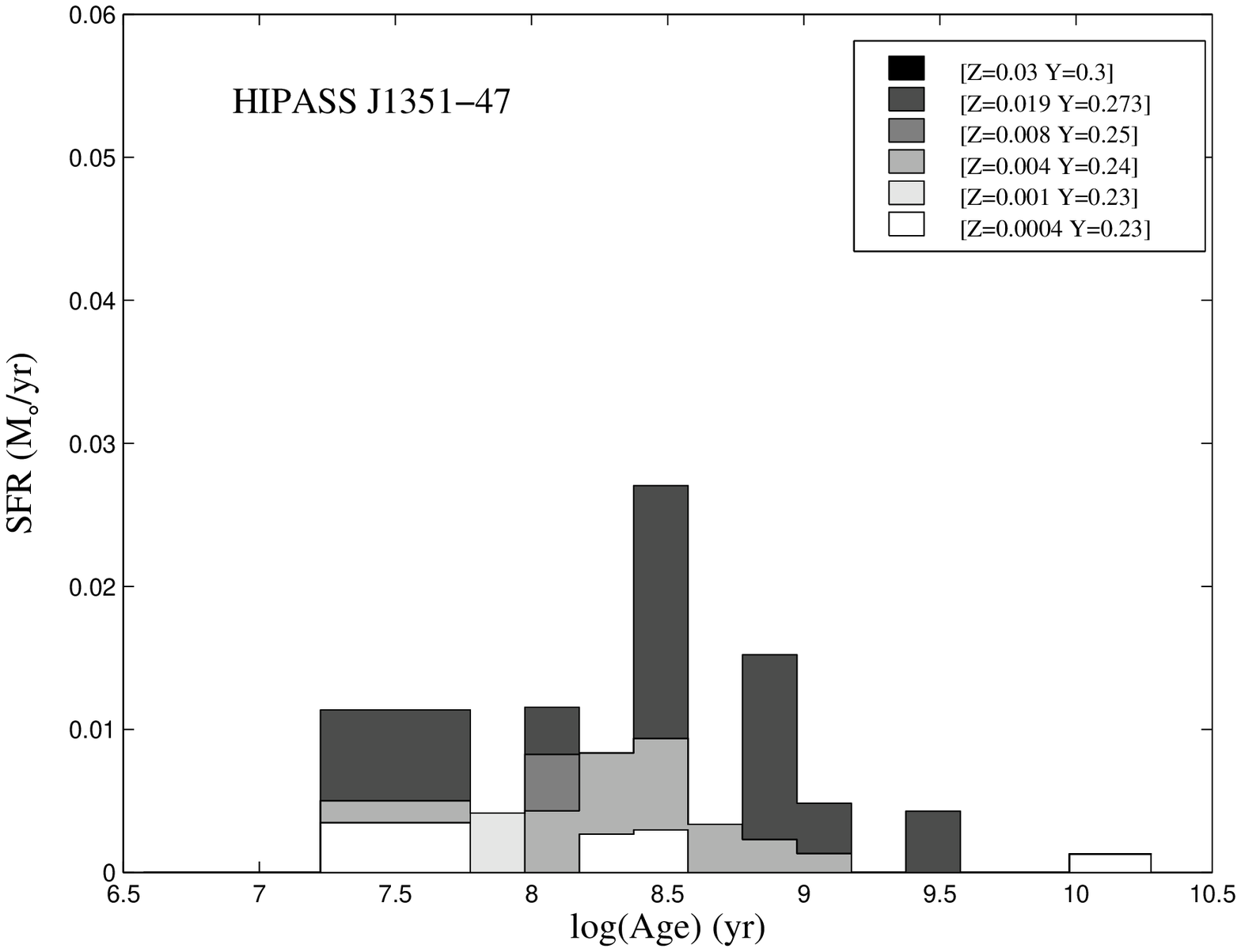}  &
\includegraphics[width=0.5\textwidth]{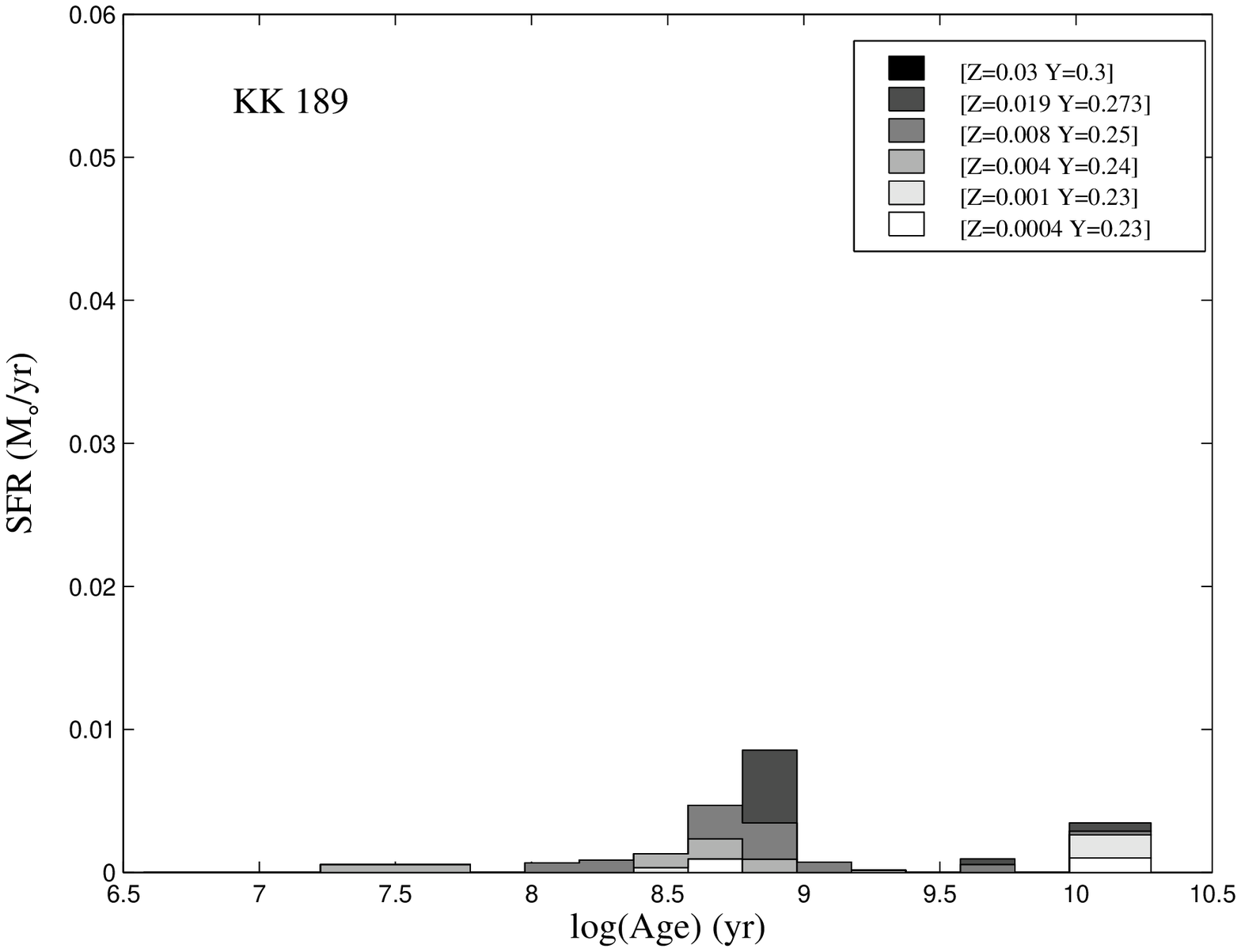}  \\
\includegraphics[width=0.5\textwidth]{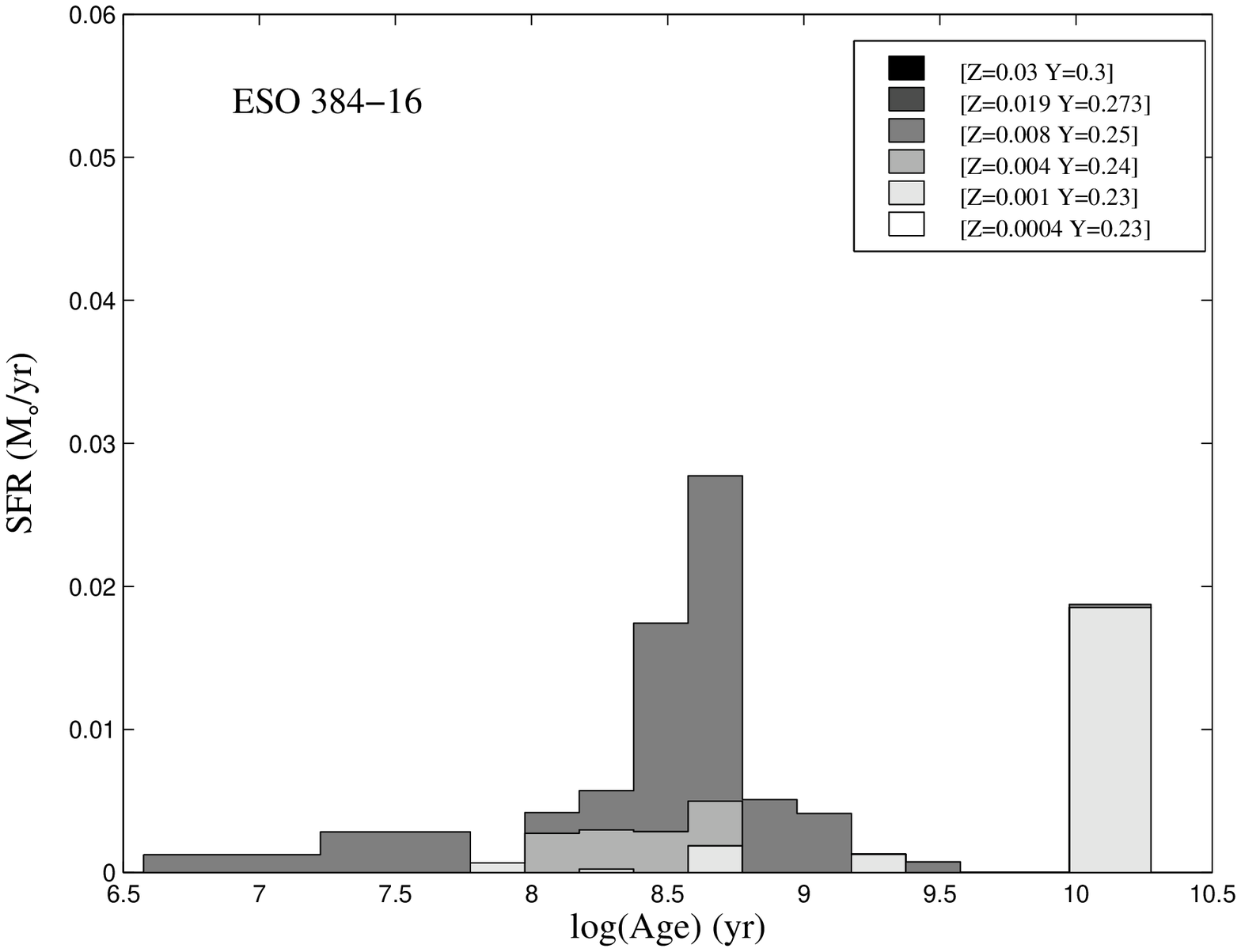}  &
\includegraphics[width=0.5\textwidth]{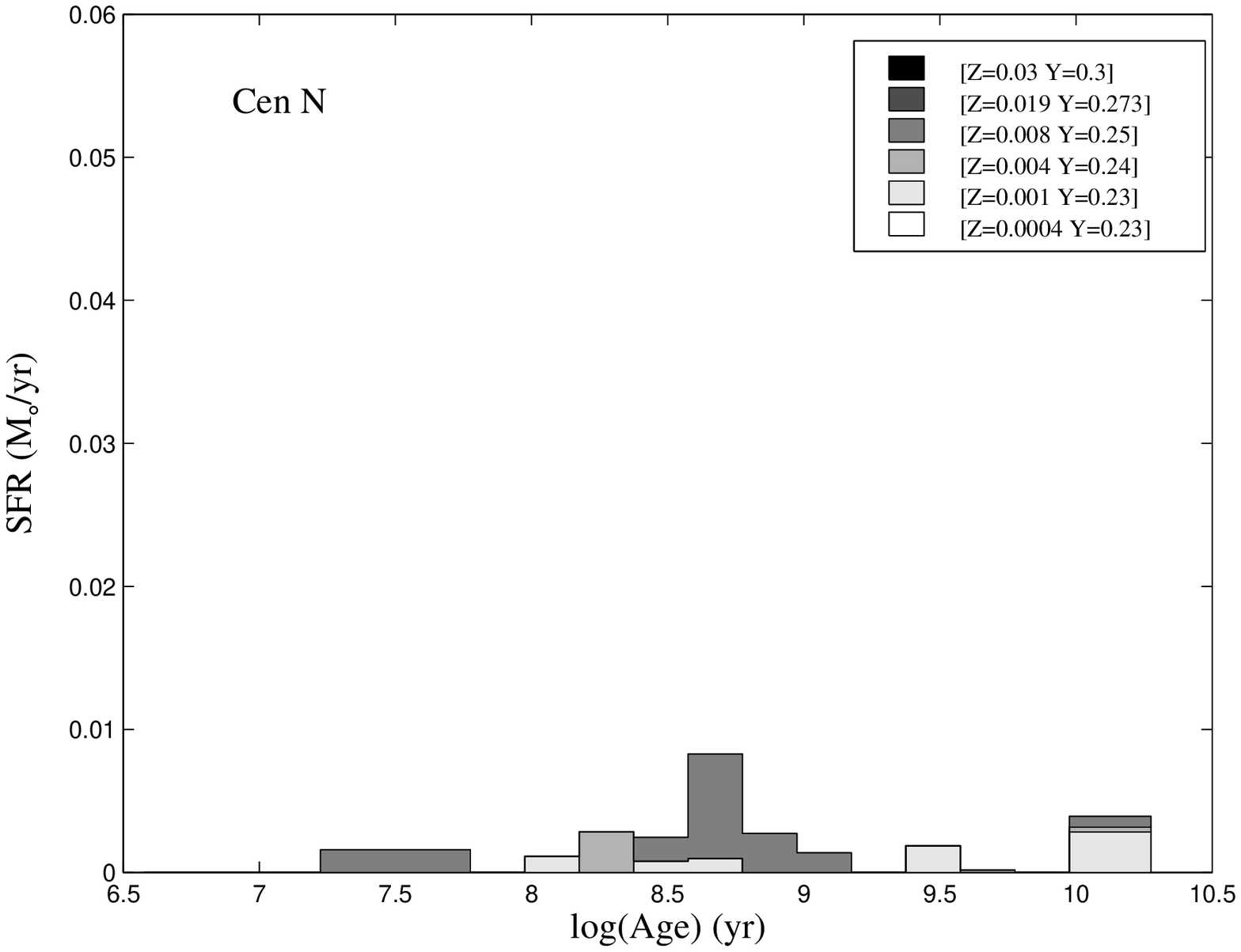}  \\
\includegraphics[width=0.5\textwidth]{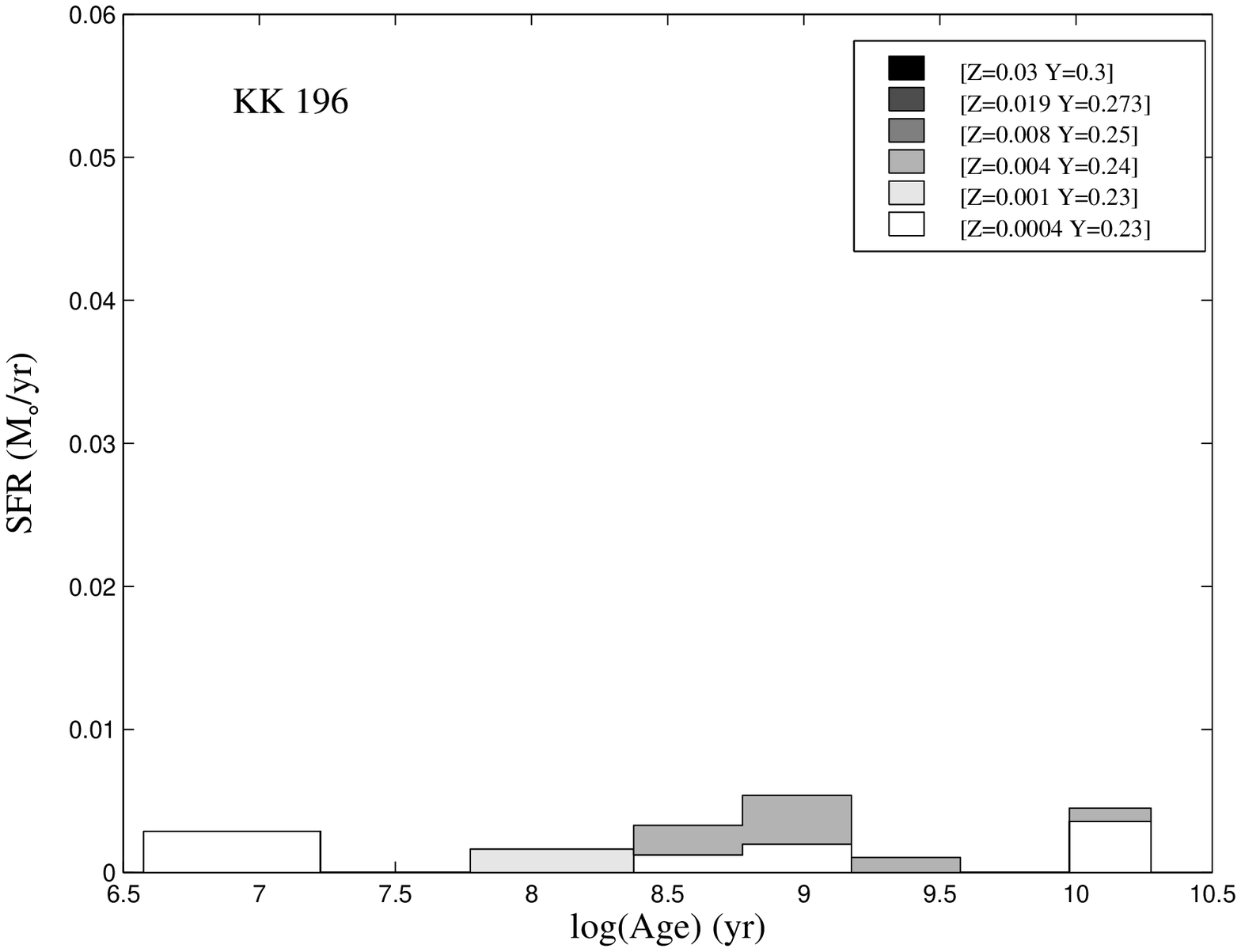}  &
\includegraphics[width=0.5\textwidth]{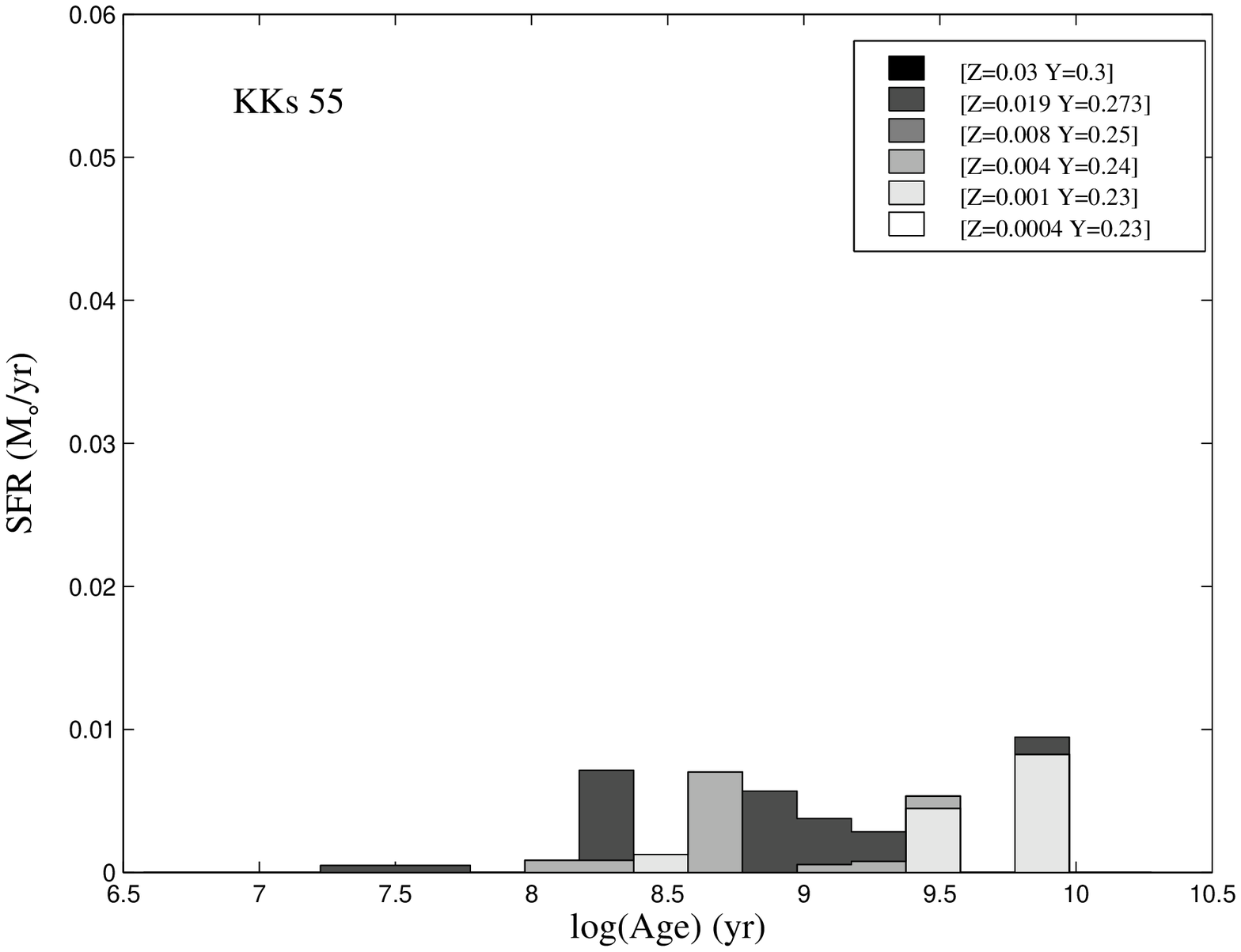} 
\end{tabular}
\caption{Star formation histories of dwarf galaxies in the Cen~A group.
Dwarf irregulars are shown at the left panels and dwarf spheroidal are at the right panels.}
\end{figure}

Several star formation features can be recognized from our measurements (see representative set in Fig.3). First, most galaxies demonstrate two distinct (and similar) star formation episodes. An ancient episode has an age of about 10--13 Gyr and low metal abundance. The galaxies under study also show enhanced star formation about 0.1--10 Gyr ago. Dwarf irregulars has some ongoing star formation, too. Second, several dwarfs under study have unusual rate of metal enrichment, with the metallicity level of recent star formation episodes of about solar and probably higher (like KK~197 and ESO~269--066). Possible explanation of this high enrichment level can be due to interaction and merging of some galaxies during the evolution of the Cen~A group. Sales et al. (astro-ph/0706.2009v3, Millennium simulations) have demonstrated high matter spread (up to 2 virial radii) in such evolutionary processes. Smaller members in the group can experience enhanced star formation due to these interactions and merging using the processed in larger galaxies matter of higher metal abundance.

\begin{acknowledgements}
This work is supported by DFG-RFBR grant 06--02--04017 and RFBR grant 07--02--00005.
We are thankful to International Astronomical Union for providing financial support for attending IAU Symposium 244.
\end{acknowledgements}

\end{document}